\documentclass[11pt]{article}

\newcommand{\be}{\begin{equation}}
\newcommand{\ee}{\end{equation}}
\usepackage{amsmath}
\usepackage{amssymb}
\usepackage{latexsym}
\usepackage{epsfig}

\newcommand{\half}{{\textstyle {1\over 2}}}

\newcommand{\Z}{\mathbb{Z}}
\newcommand{\R}{\mathbb{R}}

\newcommand{\e}{{\check e}}
\newcommand{\x}{\Sigma}

\newcommand{\ti}{\tilde}
\newcommand{\pa}{\partial}

\newcommand{\LL}{{\cal L}}

\newcommand{\RRR}{{\hbox{\rm R\kern-2.35mm R}}}

\def\ZZZ{{\hbox{ Z\kern-1.6mm Z}}}

\newcommand{\sectiono}[1]{\section{#1}\setcounter{equation}{0}}

\def\a{\alpha}
\def\b{\beta}

\def\d{\delta}

\def\i{\iota}

\def\k{\kappa}                    

\def\s{\sigma}                                   

\def\z{\zeta}
\def\D{\Delta}

\def\G{\Gamma}

\def\P{\Pi}

\def\cL{{\cal L}} 

\begin{document}

\begin{titlepage}
\rightline{\today} 
\rightline{\tt arXiv:0908.1792}
\rightline{\tt  Imperial-TP-2009-CH-04}
\rightline{\tt MIT-CTP-4054}    
\begin{center}
\vskip 2.5cm
{\Large \bf {
The Gauge Algebra of Double Field Theory and  Courant Brackets
}}\\
\vskip 1.0cm
{\large {Chris Hull${}^1$ and Barton Zwiebach${}^2$}}
\vskip 1.0cm
{\it {${}^1$The Blackett Laboratory}}\\
{\it {Imperial College London}}\\
{\it {Prince Consort Road, London SW7 @AZ, U.K.}}\\
c.hull@imperial.ac.uk
\vskip 0.5cm
{\it {${}^2$Center for Theoretical Physics}}\\
{\it {Massachusetts Institute of Technology}}\\
{\it {Cambridge, MA 02139, USA}}\\
zwiebach@mit.edu
\vskip 1.5cm
{\bf Abstract}
\end{center}

\vskip 0.5cm

\noindent
\begin{narrower}  
We investigate the symmetry algebra of the recently 
proposed field theory  on a doubled torus  
that  describes  closed string modes 
on a torus with both momentum and winding. 
The gauge parameters are constrained fields on the doubled 
space and transform as vectors under T-duality.
The gauge algebra defines a T-duality covariant bracket.
For the case in which the parameters and fields are  
T-dual to ones that have momentum but no winding,  
we find the gauge transformations  
to all orders and show that the gauge algebra
reduces to one obtained by Siegel.  We show that the bracket for such restricted parameters is the Courant bracket. We explain how these algebras are  realised as symmetries despite the failure of the Jacobi identity.

\end{narrower}

\end{titlepage}

\newpage

\tableofcontents
\baselineskip=16pt
\section{Introduction and main results}

  The closed string on a $D$-dimensional torus background 
can be formulated in terms of an infinite set of fields that 
are in fact fields 
on the doubled torus parameterized by $D$  
spacetime coordinates $x^i$  and $D$  
additional coordinates $\tilde x_i$ dual to 
winding.\footnote{  
As explained in later sections, our notation 
covers the case in which there are a number   of non-compact dimensions.
The absence of winding in the non-compact directions requires that all fields are independent of the corresponding $\ti x$'s. For simplicity, in the introduction we focus on the case with all dimensions compact.}
In a recent paper~\cite{Hull:2009mi} we began a detailed investigation 
of a {\em double} field theory.
We focused on the  sector of closed string theory consisting of  
 fields $e_{ij} (x, \tilde x)\equiv h_{ij}(x, \tilde x) 
+ b_{ij} (x, \tilde x)$ and the dilaton $d(x, \tilde x)$. 
Here $h_{ij}$ and
$b_{ij}$ are $D\times D$ matrices depending on the 
$2D$ coordinates 
$(x^i,\ti x_i)$ and they represent
doubled gravity and antisymmetric tensor fluctuations around  
constant backgrounds $E_{ij} \equiv G_{ij} + B_{ij}$.
A T-duality
invariant, gauge invariant double field theory was constructed to cubic order in the fields.  A full construction to all orders 
remains a major challenge; if achieved, the resulting theory would likely be a consistent truncation of the complete
closed string theory. 
Our construction relied on the formulation~\cite{Kugo:1992md} of closed
string field theory on tori. Earlier work on 
double field theory includes
that of Tseytlin~\cite{Tseytlin:1990nb} and that 
of Siegel~\cite{Siegel:1993th}.

A key element in the construction of~\cite{Hull:2009mi} was the
constraint that all fields and gauge parameters must be annihilated
by the operator $\Delta$
 given by
\be
\label{deltaconstraint}
\Delta = -\,2  \, {\partial\over \partial x^i}  {\partial \over \partial \tilde x_i}=  -\,2\,  {\partial_i}  {\tilde \partial^i} \, , 
\ee
having set $\alpha'=1$.  
The constraint $\Delta =0$  is needed for gauge invariance and  consistency. 
It is the field theory version of the constraint $L_0 - \bar L_0=0$ 
of closed string field theory~\cite{Zwiebach:1992ie}.   
 Since $\Delta$ is a second-order differential operator, the product of two fields in the kernel of $\Delta$ need not be in the kernel of $\Delta$.
This means that a projection
to the kernel of $\Delta$ is necessary   in the products that appear in the action and gauge transformations of the double field theory~\cite{Hull:2009mi}.
These projectors and certain cocycle factors make the double field theory non-standard and novel.  

The theory simplifies considerably if the fields and gauge parameters are restricted to be independent
of $\ti x_i$, as the constraint  $\Delta=0$  is then automatically satisfied. No projectors or cocycles are needed and it reduces to a conventional field theory of fields depending only on $x^i$.  
Similarly, a conventional field theory arises for fields and parameters that depend just on $\ti x_i$, or 
just on the coordinates of any $D$-dimensional  toroidal 
subspace $N$  
of the double space that is totally null
with respect to the 
signature $(D, D)$ metric
\be 
ds^2= 2dx^i d\ti x_i \,.  
\ee  
T-duality transforms any such  totally null or maximally isotropic subspace into 
another.
We will refer to fields with dependence only on a null subspace as {\it restricted} fields, and in so doing we will always require all fields and parameters to be restricted to the 
{\it same} space $N$.
 Any fields constructed with products
of restricted fields are 
automatically restricted.  
Restricted
fields and their products satisfy $\Delta =0$.  The converse is not true:
a  set of fields in the kernel of $\Delta$ need not 
be  restricted because
products of the fields need not be in the kernel of $\Delta$. The reduction to restricted fields  -- discussed in detail in \S\ref{restridual}  --  is thus a significant truncation of  
the field space of double field theory.
  
\medskip

String theory on a $D$-torus has an $O(D,D;\Z)$ T-duality symmetry 
and 
double field theory  
inherits 
this T-duality symmetry.  
In fact, the double field theory would 
have a formal  $O(D,D)$ symmetry if all dimensions were non-compact. We formulate  
double field theory in a way that is largely independent of the number of dimensions that are toroidal, and we will find it convenient to refer to expressions as being $O(D,D)$ covariant (or invariant) when we in fact mean covariant (or invariant) under
 the subgroup of $O(D,D)$ preserving the boundary conditions on 
 periodic coordinates,  together with the condition of 
 no winding in non-compact directions.
A central role will be played  by the requirement that the  gauge algebra 
 have such $O(D,D)$ covariance.

In this paper we focus on the gauge transformations and their algebra.
 We find here the complete non-linear transformations for restricted fields and the corresponding gauge algebra. 
 Our results are   $O(D,D)$ covariant, so that they apply for any choice of null subspace for the restricted fields.
  The constraint to restricted fields was also assumed by 
  Siegel~\cite{Siegel:1993th}.    
  Using  a set of fields larger 
than the one used here, 
as well as additional gauge invariances, he proposed
a gauge and duality invariant action to all orders in the fields.  Most of our
work in this paper relates to restricted fields, so comparison with the
 results of~\cite{Siegel:1993th}  
will feature at various points. In particular, we show that our gauge algebra reduces to that of Siegel when the gauge parameters are restricted.

\medskip
To first order in the fields, the gauge
transformations of  $e_{ij} (x, \tilde x)$ and   $d(x, \tilde x)$
take the form~\cite{Hull:2009mi}
\be
\label{kssmc}
\begin{split}
\delta_\lambda
 e_{ij}  &= ~   D_i \bar\lambda_j + \bar D_j \lambda_i \, 
+   {1\over 2}\,(\lambda \cdot D + \bar \lambda \cdot \bar D) e_{ij} 
+{1\over 2} \,   (D_i \lambda^k - \,  D^k \lambda_i)\, e_{kj}    
    \, - e_{ik} \,\, {1\over 2} 
   ( \bar D^k \bar \lambda_j - \bar D_j \bar \lambda^k )\,, \\[1.0ex]
\delta_\lambda  d~  &= - {1\over 4}  (D\cdot \lambda +\bar D\cdot \bar \lambda )
+  {1\over 2}  (\lambda \cdot D+\bar\lambda \cdot \bar D) \,d\,~.
\end{split}
\ee
Here $\lambda_i(x, \tilde x)$ and $\bar \lambda_i(x, \tilde x)$ are independent  real gauge parameters, 
and the derivatives $D, \bar D$ are defined~by
\be
\label{gro8774}
D_i =\,{\partial_i} - E_{ik} \,{\tilde\partial ^k}\,, \qquad
\bar D_i = \,{\partial_i} + {E}_{ki}\, {\tilde\partial^k}\,.
\ee
 They are independent real derivatives 
with respect 
to right- and left-moving coordinates~\cite{Hull:2009mi}. 
Indices are  raised and lowered with the background metric $G_{ij}$ and $a\cdot b = G^{ij} a_i b_j$. It is straightforward to verify that $D^2 - \bar D^2 = 2\Delta$.   
The above  gauge transformations are reducible. If we take
\be
\label{reducible}
\lambda _i = D_i \chi, \qquad \bar  \lambda _i = -\bar D_i \chi\,,
\ee
for arbitrary   
$\chi(x, \ti x)$, then the fields are left invariant.  Therefore,  
$\lambda _i\to \lambda _i+ D_i \chi $ and $ \bar  \lambda _i \to  \bar  \lambda _i  -\bar D_i \chi$ constitutes a ``symmetry for a symmetry." This will play an important role in our discussion.

The gauge algebra is $[\delta_{\lambda_1}\,, 
\delta_{\lambda_2} ] = \delta_{\lambda_{12}} + \cdots$, where to leading nontrivial order
\be
\label{ixltfttasvm}
\begin{split}
\phantom{\Biggl[}~~\lambda^i   _{12}=
&\,  ~~{1\over 2} \, (\lambda_2 \cdot D + \bar \lambda_2 \cdot \bar D) \, \lambda_1^i 
-{1\over 4} \,
\Bigl[ ~ \lambda_2 \cdot D^i  \lambda_1 - \bar \lambda_2 \cdot D^i  \bar \lambda_1 \Bigr] \,-\, (1\leftrightarrow 2) \,,
\\[1.3ex]
\bar \lambda^i _{12}= & 
~~{1\over 2} \,  (\lambda_2 \cdot D + \bar \lambda_2 \cdot \bar D)\, \bar\lambda_1^i 
+ {1\over 4} \,\Bigl[ ~ \lambda_2 \cdot \bar D^i  \lambda_1 - \bar\lambda_2 \cdot \bar D^i  \bar \lambda_1  \Bigr]\,-\, (1\leftrightarrow 2)~.
\end{split}
\ee
A projection 
to the kernel of $\Delta$ is necessary (and implicit) in the products that appear in (\ref{kssmc})
and in (\ref{ixltfttasvm}). 
For restricted fields and parameters,
however, the projections are not needed
as the $\D=0$ constraint is automatically satisfied.
The gauge algebra (\ref{ixltfttasvm}) defines a bracket of two gauge parameters and this
bracket   does not satisfy the Jacobi identity~\cite{Hull:2009mi}.

\medskip
 
 The results above  for the gauge transformations and algebra respect certain rules
of index contraction that arise from the string theory and result in
$O(D,D)$ or T-duality covariance of the double field theory.  While we display just
one kind of index $i, j, k, \ldots$, some indices should be thought
as barred and some as unbarred.  On the field $e_{ij}$ the first index, $i$, is unbarred
and the second, $j$, is barred.  On $\bar\lambda_i$ and $\bar D_i$ the index
is barred.  On $\lambda_i$ and $D_i$ the index is unbarred. Contractions
can only occur between indices of the same type, and equations
must relate objects with identical index structure.  
The metric $G_{ij}$ used to contract indices can be viewed as having two barred or two unbarred indices.\footnote{These rules of contraction can be understood using 
$O(D)_L\times O(D)_R$ indices, with unbarred indices for $O(D)_L$ and barred ones for $O(D)_R$. }

We can ask if the gauge transformations and gauge algebra displayed
above can be extended to all orders for restricted fields.
We answer this question in the affirmative in 
\S\ref{thegaugetransformations}. We show that the only non-linear correction  is the addition to $\delta_\lambda e_{ij}$ of a term quadratic
in $e_{ij}$.  The result is given in (\ref{finalgt}). 
 The gauge transformations then close off-shell for restricted fields.  In fact, the full algebra
remains that of (\ref{ixltfttasvm}) with (field independent) structure 
constants.
The full gauge transformations remain reducible, 
 and  
  parameters of the form (\ref{reducible}) still leave the fields invariant.
We show that our gauge transformations are related to the standard ones for a metric and $B$-field via field redefinitions.
We stress that the algebra closes without use of the equations of motion, but with repeated use of the condition that the fields and parameters are restricted.
 
To clarify the meaning of the gauge transformations,  in \S\ref{exploring}  we rewrite the
gauge parameters $\lambda_i, \bar \lambda_i$ in terms of 
quantities $\xi^i, \,\tilde\xi_i$,  
and use $O(D,D)$ 
covariant notation, defining
 \be
 \label{etis}
\partial _M\equiv    \begin{pmatrix}\,\ti \partial ^i\, \\[1.0ex] 
 \partial _i
 \end{pmatrix}\,  ~~~~
 \x^M \equiv    \begin{pmatrix} \,\ti \xi _i\, \\[1.0ex] 
\xi^i \end{pmatrix}\,, ~~\eta_{MN}  =\begin{pmatrix}0& I \\ I& 0 \end{pmatrix} 
\,.
\ee
The familiar transformations arise for parameters that are independent of $\ti x$, so that
 the gauge parameters $\xi^i(x)$ and $\ti \xi_i(x)$ are a vector field and a one-form over the spacetime $M$ with coordinates $x^i$. These are  related to the parameters for infinitesimal diffeomorphisms of $M$ and 
 $B$-field
  gauge transformations respectively~\cite{Hull:2009mi}.

The gauge 
algebra (\ref{ixltfttasvm}) for general $\x( x,\ti x)$ 
can be rewritten as 
$[\delta_{\x_1}\,, 
\delta_{\x_2} ] = \delta_{\x_{12}} $
where
$\x_{12}  =- [\x_1, \x_2]_C$
and the C-bracket is defined by
\be
\label{algfinalform9}
 [\x_1, \x_2]_{{}_C}  \equiv  ~ 
  \x^N_{[1} \partial_N \, \x^M_{2\hskip1pt]}
-  \,{1\over 2}  \,
 \eta  ^{MN }\eta _{PQ} \,\x_{[1}^P \, \partial_N \,  \x_{2]}^Q\,\,.
 \phantom{\Biggl[}  
\ee

\smallskip
\noindent
We use the convention $U_{[r} V_{s]} =  U_r V_s - U_s V_r$.
The first term in the 
right-hand side  of  (\ref{algfinalform9})
 is the Lie bracket $ [\x_1, \x_2] =  
\x^N_{[1} \partial_N \, \x^M_{2]}$ while the second is unexpected.
In this form, the gauge algebra coincides with that derived
by Siegel~\cite{Siegel:1993th}
 (where the derivative defined using the C-bracket was called a `new' Lie derivative). 
  The C-bracket is manifestly $O(D,D)$ covariant.   
With $O(D,D)$  notation, 
 the transformations (\ref{reducible}) that do not act on the fields have parameters that take the form    
 \be 
 \label{reduciblecov}
 \x^M= \eta ^{MN} \partial _N \chi\,.
 \ee

Remarkably, the C-bracket that arises in the gauge algebra here is related to brackets that have been prominent  in the mathematics literature.
In \S\ref{courantbrackets} we show that for parameters 
restricted to be independent of 
$\ti x_i$, 
the C-bracket is precisely the Courant bracket~\cite{Tcourant},  
a central construction in generalised  geometry~\cite{Hitchin,
Gualtieri,LWX}.
Indeed, 
  $\ti x$-independent gauge parameters $\xi^i(x)$ and $\ti \xi_i(x)$ 
together give a section of  the formal sum $(T\oplus T^*)M$ of tangent and cotangent bundles
and the
 Courant bracket is defined on such sections.  
Parameters restricted to 
an arbitrary null $N$ can be regarded as sections of $(T\oplus T^*)N$ and the C-bracket becomes  the Courant bracket on $(T\oplus T^*)N$. 
The choice  of $N$   breaks the $O(D,D)$  covariance of  the C-bracket.
Since the choice of $N$ need not be made explicit, 
the C-bracket can be regarded as an $O(D,D)$ covariantization of the Courant bracket.
In \S\ref{cbracksec}   we show that it is  equivalent (for restricted parameters) to the Courant-like bracket of~\cite{LWX}
that treats vectors and one-forms 
symmetrically. 

Neither the C-bracket nor the Courant bracket   
satisfy the Jacobi identity.
It is then natural to ask  
how  
this failure of the Jacobi identity
can be consistent with the
realisation of these brackets
 in a symmetry algebra.
To answer this question  we consider the
associated infinitesimal field transformations
$\delta_\Sigma$. 
The commutator of two transformations acting on fields gives  
\be
\label{eorcfgkj}
[\delta_{\Sigma_1}, \delta_{\Sigma_2} ] = \delta_{[\Sigma_1, \Sigma_2]} \,. 
\ee
Here $[\Sigma_1 \,, \Sigma_2]$  is the bracket of gauge parameters, which for our case is ($-1$ times) the C-bracket.
The bracket has a non-vanishing Jacobiator $J$,   
 defined by
\be
\label{jacobiator-def}
  {J(\Sigma_1, \Sigma_2, \Sigma_3)} \equiv   
 {[\, [{\Sigma_1} , {\Sigma_2}]\,, {\Sigma_3} \, ] 
+ [\, [{\Sigma_2} , {\Sigma_3}]\,, {\Sigma_1} \, ] 
+ [\, [{\Sigma_3} , {\Sigma_1}]\,, {\Sigma_2} \, ] }     \,.
\ee
The commutators of transformations automatically satisfy
\be
 [\, [\delta_{\Sigma_1} , \delta_{\Sigma_2}]\,, \delta_{\Sigma_3} \, ] 
+ [\, [\delta_{\Sigma_2} , \delta_{\Sigma_3}]\,, \delta_{\Sigma_1} \, ] 
+ [\, [\delta_{\Sigma_3} , \delta_{\Sigma_1}]\,, \delta_{\Sigma_2} \, ] 
= 0 \, ,
\ee
when acting on fields.
The left hand side, however,   
 can be evaluated using (\ref{eorcfgkj}) 
to give $\delta_{J(\Sigma_1, \Sigma_2, \Sigma_3)}  $
which can only be consistent with the above condition if
$\delta_{J(\Sigma_1, \Sigma_2, \Sigma_3)}  $
 is zero when acting on fields.
This requires that the Jacobiator $J(\Sigma_1, \Sigma_2, \Sigma_3)$ be a parameter of the form (\ref{reduciblecov}) for a trivial gauge
transformation leaving all fields invariant. As we will discuss, the Jacobiators for the C-bracket and the Courant bracket are precisely of this form, so that the algebra can indeed be consistently realised on fields.

In a theory with redundant gauge symmetry,
the gauge algebra is ambiguous since the bracket of two gauge
parameters can be changed by adding a parameter that generates
a trivial symmetry.
 If $\s_{12}$ is any 
nonvanishing gauge parameter of the form (\ref{reduciblecov}) so
that $\d_{\s_{12}}$ leaves all fields invariant, then  (\ref{eorcfgkj}) can be changed
to 
\be
\label{eorcfgkj99}
[\delta_{\Sigma_1}, \delta_{\Sigma_2} ] = \delta_{[\Sigma_1, \Sigma_2] + \s_{12}} \,. 
\ee
We shall argue that no such $ \sigma_{12}$    
can be constructed from $\Sigma_1$ and $ \Sigma_2$
in a way that is $O(D,D)$ covariant. It is an important 
point in this paper that the duality
symmetry $O(D,D)$ fixes this ambiguity and the C-bracket cannot
be changed while preserving  $O(D,D)$ covariance.
 This ambiguity, however,   
  plays a useful role in relating our results to those for conventional field theory, which are not $O(D,D)$ covariant.
The Jacobiator and redundant symmetries will be discussed further
 in~\S\ref{Jacalg}.

\sectiono{The gauge transformations}\label{thegaugetransformations}

\medskip

Our aim in this section is to   investigate the higher order corrections to the gauge transformations and algebra reviewed in the introduction.  
We assume  fields and gauge parameters restricted to some isotropic
subspace $N$.  For such fields  no projection
to the kernel of $\Delta$ is needed and the cocycles vanish, so that the calculations are those for a conventional classical field theory.
We work in an $O(D,D)$ covariant framework so that the result applies for {\it any} choice of isotropic subspace $N$. We 
find the full non-linear transformations and algebra for restricted fields.

Before proceeding, we briefly discuss our notation; see~\cite{Hull:2009mi} for further details.
The simplest case is that in which all $D$ coordinates $x^i$ are periodic, but our notation also 
covers the case in which there are $n$ non-compact dimensions.
Then $x^i=(x^\mu, x^a)$ split into coordinates   
  $x^\mu$ on the $n$-dimensional Minkoswski space $\R^{n-1,1}$  and  coordinates  $x^a$ on the $d$-torus $T^d$ where $n+d=D$.
  The absence of winding in the $x^\mu$ directions requires that all fields and gauge parameters are independent of $\ti x_\mu$
  so that $\pa/ \pa \ti x _\mu =0$  on all fields and   parameters.
The fields then depend only on $x^\mu, x^a, \ti x_a$. T-duality is the group  
$O(d,d;\Z)$ acting on the doubled torus with coordinates $x^a,\ti x_a$. 
For restricted fields, the totally null subspace $N$   has coordinates
$x^\mu$ together with a totally null $d$-dimensional torus subspace of the torus $( x^a,\ti x_a)$, e.g.
  $(x^\mu, x^a)$ or $(x^\mu, \ti x_a)$. 
Such spaces $N$ are related by the action of $O(d,d;\Z)$.

From~\cite{Hull:2009mi}  and the discussion in the introduction, the full form of the gauge transformations should have the following properties for restricted fields:   
 \begin{itemize}

\item   The gauge algebra closes.

\item  The transformations are $O(D,D)$ covariant.

\item    All index contractions in the transformations should be of the allowed kind of barred indices with barred indices or of unbarred indices with unbarred indices. 
This is necessary for  $O(D,D)$ covariance.
\item  
For any choice of isotropic subspace $N$, the  transformations should be related to the  standard Einstein plus 
$B$-field transformations by redefinitions of the fields and parameters.

\end{itemize}

We start by investigating the last two criteria. 
These will be sufficient to find the full non-linear form of the gauge transformations  for restricted fields and 
we will then check the algebra of these transformations in an appendix.
Consider the restriction to  fields that have no $\tilde x$ dependence, 
so that  setting $D = \bar D = \partial$ 
brings the transformations  (\ref{kssmc}) to a form that can be related to the standard gauge transformations.  
In the standard Einstein plus 
$B$-field theory, the gauge transformations  are the diffeomorphisms 
with infinitesimal parameter $  \epsilon ^i$
and antisymmetric tensor gauge transformations with infinitesimal parameter $ \tilde \epsilon_i$.
The full  metric is $G_{ij}+h_{ij}$ and the  
antisymmetric tensor gauge field is $B_{ij}+b_{ij}$ where $G_{ij}$ and $B_{ij}$ are constant background fields.
For the combined fluctuation field
$\e_{ij}=h_{ij} +b_{ij}$, the transformations are
\be
\delta \e_{ij} =  \delta^{(0)}\e_{ij}  +  \delta^{(1)}\e_{ij}\,,
\ee
with
\be
\label{reovk}
\begin{split}
 \delta^{(0)}\e_{ij} &= \partial_i \epsilon_j + \partial_j \epsilon_i
 - (\partial_i \tilde \epsilon_j - \partial_j \tilde \epsilon_i) \,, \\[0.6ex]
  \delta^{(1)}\e_{ij} &= \epsilon^k \partial_k  \e_{ij} + (\partial_i \epsilon^k)
  \,\e_{kj}  +  \e_{ik} \,(\partial_j \epsilon^k) \,.
\end{split}
\ee
 Here and in what follows, indices $i,j$ are raised and lowered using the background metric $G_{ij}$.

In order to connect with our formalism we rewrite $\epsilon$ and $\bar \epsilon$ in terms of $\lambda$ and $\bar\lambda$:
\be
\label{lamdis}
\epsilon_i =  {1\over 2} (\lambda_i + \bar \lambda_i)\,, ~~
\tilde\epsilon_i =  {1\over 2} (\lambda_i - \bar \lambda_i)\,.
\ee
 In~\cite{Hull:2009mi} it was shown that  
  the field $ \e_{ij}$ is related to $ (e_{ij},d)$ by  
  $ \e_{ij} = 
  f_{ij}(e,d)$, where
\be
\label{fullfieldbx}
  f_{ij}(e,d) =e_{ij}  +  {1\over 2} {e_i}^{\,k} e_{kj}  
 + \hbox{cubic corrections}\,, 
 \ee 
 and this maps the transformations (\ref{reovk}) to (\ref{kssmc}),
  up to terms quadratic in fields.
 The redefinition takes the full non-linear transformations 
 of $\e_{ij}$ 
  given in (\ref{reovk}) to transformations of
 $e_{ij}$, and the condition that there should only be allowed  contractions in the transformation of $e$ places stringent constraints on~$f$. 
The redefinition
 $ \e_{ij} = 
  f_{ij}(e,d)$
gives terms quadratic in the fields that include ones with disallowed contractions. These \lq bad terms'    can be eliminated by taking 
\be
\label{fullfieldb}
 f_{ij}(e,d) =e_{ij}  +  {1\over 2} \,{e_i}^{\,k} e_{kj}  + {1\over 4} {e_i}^{\,k} e_{kl}\,e^l{}_j
 + \hbox{quartic corrections}\, .
 \ee 
 This is easily extended to arbitrary order, and one soon finds that requiring only allowed contractions fixes $f$ to be
\be
\label{mich1}
f =  \Bigl( 1- {1\over 2} \,e\Bigr)^{-1} e  \,,   
\ee
so that
\be
\label{mich11}
\e =  \Bigl( 1- {1\over 2} \,e\Bigr)^{-1} e  \,,   
\ee
where we use matrix notation, so that the first few terms are as in (\ref{fullfieldb}). 
The function (\ref{mich1}) first arose in the work of
Michishita~\cite{Michishita:2006dr}.\footnote{
This function was proposed in~\cite{Michishita:2006dr} as a simple ansatz 
for  the relation between the conventional variable $\e$ and the string field  variable 
that satisfies a number of constraints but is not uniquely selected. 
It was
shown in~\cite{Hull:2009mi} \S4.5,  that the 
relation between $\e$ and the string field variable has explicit dependence on $d$.}
We now show that this gives no bad contractions and use this to find the full non-linear gauge transformation of $e_{ij}$.

It is an immediate consequence of the definition (\ref{mich1})  that $\e$ and $e$
commute:
\be
\label{mich2}
\e \, e =  e \, \e \,.
\ee
It follows 
that
\be
\label{mich3}
e  = \Bigl( 1- {1\over 2} \, e \Bigr) \,\e  = \e\, \Bigl( 1- {1\over 2} \, e \Bigr) \,.
\ee
The above leads to
\be
\label{mich4}
\e - e = {1\over 2} \, e\e  =  {1\over 2}  \e e \,,
\ee
and one readily verifies that
\be
\label{mich5}
\Bigl( 1+ {1\over 2} \, \e \Bigr) \Bigl( 1- {1\over 2} \, e \Bigr)  = 1\,.
\ee
Finally, varying~(\ref{mich4}) and using   (\ref{mich5}) we find
a relation between arbitrary variations,
\be
\label{mich6}
\delta e   =   \Bigl( 1- {1\over 2} \, e \Bigr) \, \delta \e \,   \Bigl( 1- {1\over 2} \, e \Bigr)\,.
\ee

The standard gauge
transformations (\ref{reovk}) can be rearranged using (\ref{lamdis}) to give
\be
\label{reovk1}
\begin{split}
 \delta^{(0)}\e_{ij} &= \partial_i \bar \lambda_j + \partial_j \lambda_i\,,  
 \\[1.0ex]
  \delta^{(1)}\e_{ij} &= {1\over 2}  \Bigl[ (\lambda + \bar\lambda)\cdot \partial \Bigr]
   \e_{ij} + \Bigl[ {1\over 2}\bigl(  \delta^{(0)}\e_i^{~k}\bigr)  + {\cal N}_i^{~k}\Bigr]
  \,\e_{kj}  +  \e_{ik} \, \Bigl[ {1\over 2}\bigl(  \delta^{(0)}\e^k_{~j}\bigr)  - 
  \bar{\cal N}^k_{~j}\Bigr]\,,
\end{split}
\ee
where 
\be
\begin{split}
 {\cal N}_i^{~k} &= \partial_i \lambda^k - \partial^k \lambda_i \,,\\[0.3ex]
  \bar{\cal N}^k_{~j} \,&= \partial^k \bar \lambda_j - \partial_j
  \bar\lambda^k \,.
\end{split}
\ee
Note that  $ \delta^{(0)}\e_{ij} =  \delta^{(0)}e_{ij} $ where 
$\delta^{(0)}e_{ij}$    is the zeroth order transformation of $e_{ij}$ in~(\ref{kssmc}).   
Using matrix notation, the full gauge transformation can be written as
\be
\label{umn1}
\delta \e =   \delta^{(0)} e  +  {1\over 2}
 \Bigl[ (\lambda + \bar\lambda)\cdot \partial \Bigr]
 \e  +  \Bigl[ {1\over 2}  \delta^{(0)} e  + {\cal N}\Bigr]\e
+ \e \, \Bigl[ {1\over 2}  \delta^{(0)} e  - \bar{\cal N}\Bigr]\,.
\ee
We now determine $\delta e$ using (\ref{mich6}).
Using (\ref{mich3}) and (\ref{mich6}) we find
\be
\begin{split}
\delta e &=  \delta^{(0)} e  - {1\over 2} e\, \delta^{(0)} e -{1\over 2} (\delta^{(0)} e ) e + {1\over 4} e (\delta^{(0)} e ) e + {1\over 2} \Bigl[ (\lambda + \bar\lambda)\cdot \partial \Bigr]  e\\
&~~+ \Bigl(1-{1\over 2}\,e\Bigr)  \Bigl[ {1\over 2}  \delta^{(0)} e  + {\cal N}\Bigr]\,e  + e \,\Bigl[ {1\over 2}  \delta^{(0)} e  - \bar{\cal N}\Bigr]
\Bigl(1-{1\over 2}\,e\Bigr)\,.
\end{split}
\ee
Expanding out and cancelling some terms we find
\be
\label{erlkn}
\delta e = \delta^{(0)} e + {1\over 2}  \Bigl[ (\lambda + \bar\lambda)\cdot \partial \Bigr] e
+  {\cal N} e  - e\, \bar  {\cal N} -{1\over 4} \, e  \, (\delta^{(0)} e )^t \,e\,,
\ee
where we made use of the identity  
\be
{1\over 2} \,\delta^{(0)} e  + {\cal N} - \bar {\cal N} = {1\over 2} \,
(\delta^{(0)} e)^t\,.
\ee
Restoring explicit indices in (\ref{erlkn}) we obtain
\be
\begin{split}
\delta_\lambda
 e_{ij}  &= ~   \partial_i \bar\lambda_j + \partial_j \lambda_i \, \\[1.0ex]
&~+   {1\over 2}\,
 (\lambda ^k  + \bar \lambda ^k)   \partial  _k  \, 
 e_{ij} 
+{1\over 2} \,   (\partial_i \lambda^k - \,  \partial^k \lambda_i)\, e_{kj}    
    \, - e_{ik} \,\, {1\over 2} 
   ( \partial^k \bar \lambda_j - \partial_j \bar \lambda^k ) \\[1.0ex]
&~ - {1\over 4}\, e_{ik}\,(\partial^l \bar \lambda^k +\partial^k \lambda^l)\, e_{lj} \,.
   \end{split}
   \ee 
 This is the full non-linear transformation for fields that are independent of $\ti x$ and indeed has no bad contractions.
 For general polarisations, some of the derivatives $\partial _i$ should become $D_i$ and some should become $\bar D_i$. There is a unique way of doing this which 
 uses only 
 allowed contractions:    
\be
\label{finalgt}
\begin{split} \phantom{\Biggl(}
\delta_\lambda
e_{ij}  &= ~   D_i \bar\lambda_j + \bar D_j \lambda_i \, \\
&~+   {1\over 2}\,(\lambda \cdot D + \bar \lambda \cdot \bar D) e_{ij} 
+{1\over 2} \,   (D_i \lambda^k - \,  D^k \lambda_i)\, e_{kj}    
    \, - e_{ik} \,\, {1\over 2} 
   ( \bar D^k \bar \lambda_j - \bar D_j \bar \lambda^k )~~ \\[0.3ex]
&~ - {1\over 4}\, e_{ik}\,(D^l \bar \lambda^k +\bar D^k \lambda^l)\, e_{lj} \,.
\phantom{\Biggl(}   \end{split}
   \ee
This is our final answer for the gauge transformations.
Note that the $\delta^{(1)}$ transformations 
derived in~\cite{Hull:2009mi} and cited in the introduction  
are correctly generated. It is remarkable that only one extra term quadratic in fields is needed, so that the transformations are polynomial.
Since the gauge algebra  of the standard transformations closes, the transformations we obtained from these by field redefinitions should also have a closed algebra.
In fact they do, as we have confirmed explicitly
by direct computation (details in the Appendix), and the gauge algebra is precisely~(\ref{ixltfttasvm}).
Note that this algebra has structure constants, not field-dependent structure functions.

The gauge transformation  (\ref{kssmc}) of the dilaton
satisfies the gauge algebra~(\ref{ixltfttasvm}) and only  involves good contractions.
For restricted fields, these transformations can be obtained from 
those of the scalar dilaton $\phi$ and string-frame metric $g_{ij}$ by the field relation
$e^{-2d}= e^{-2 \phi}\sqrt {- g}$, as discussed in~\cite{Hull:2009mi}. 
Then we can take (\ref{kssmc}) as the full transformations of $d$ exact to all orders in the fields (for restricted fields). 
 This can be thought of as fixing the field-redefinition ambiguity.
It is straightforward to check that gauge   
transformations with
gauge parameters  
(\ref{reducible}) leave both $e_{ij}$ and $d$ invariant so that, as expected,
the gauge symmetry is still 
reducible.   It remains to discuss the $O(D,D)$ covariance of the transformations and algebra.
For this it is convenient
to streamline the notation, as we do next.

\sectiono{O(D,D) rewriting of the gauge algebra }\label{exploring} 

In this section we rewrite the gauge algebra~(\ref{ixltfttasvm}) 
in a  simpler form, using $O(D,D)$ covariant notation and   
the formalism introduced in~\cite{Hull:2009mi}. 
As a first step, we define 
$\xi^i$ and 
$\tilde\xi_i$ in terms of the gauge  
parameters $\lambda$ and $\bar \lambda$:
\be
\label{latoxi}
\xi ^i \equiv{ 1\over 2}( \lambda^i   +\bar  \lambda^ i), \qquad
\ti \xi _i \equiv { 1\over 2}(-E_{ji} \lambda^j  +E_{ij}\bar  \lambda^ j)\,.
\ee
For reference, we also record the inverse relations:
\be
\lambda_i = - \tilde\xi_i + E_{ij} \xi^j \,, ~~~~\bar\lambda_i  = \tilde\xi_i
+ E_{ji} \,\xi^j  \,.
\ee
In the above indices are raised and lowered with the background
metric  $G_{ij}$.  As we noted in the introduction, 
 the   partial derivatives
$(\partial_i , \tilde \partial^i)$  with respect to the
coordinates $(x^i, \tilde x_i)$ are related to 
the derivatives $(D_i, \bar D_i)$ by   
\be
\label{gro8774}
D_i =\,{\partial_i} - E_{ik} \, \tilde\partial^k\,, \qquad
\bar D_i =\,\partial_i + E_{ki}\,\tilde\partial^k  \,,
\ee
with the inverse relations
\be
    \tilde \partial^i  
= {1\over 2} (-D^i + \bar D^i )\,, ~~~~~~ \, \partial_i  = {1\over 2} ( E_{ji}D^j  +  E_{ij} \bar D^j )\,.
\ee
It is then straightforward to verify that 
\be
\label{aux01}
 \frac 1 2(\lambda^i D_i +\bar  \lambda^ i \bar D_i
 )=
\xi^i \pa_i
+ \ti \xi _i \ti \pa ^i  \,.
\ee
 
Following~\cite{Hull:2009mi}, we can combine $x$ and $\tilde x$ coordinates,  $\partial$ and $\tilde \partial$ derivatives, and 
$\xi$ and $\tilde \xi$ parameters 
into $O(D,D)$ covariant expressions
\be 
 \label{vecx} 
X ^M\equiv  \begin{pmatrix} \,\tilde x_i\, \\[0.6ex] x^i  \end{pmatrix}\,, 
~~~~\partial _M\equiv    \begin{pmatrix}\,\ti \partial ^i\, \\[1.0ex] 
 \partial _i
 \end{pmatrix}\,,  ~~~~
 \x^M \equiv    \begin{pmatrix}\, \ti \xi _i\, \\[1.0ex] 
\xi^i \end{pmatrix}\,.
\ee
Here $M=1,...,2D$.  
The original space $M$ has coordinates $x^i$ and the dual space $\ti M$ has coordinates $\ti x_i$.
Together these combine to form the doubled space 
$\hat M= M\times \ti M$  with coordinates $X^M$. 
The parameters 
$\xi^i (x,\ti x)$ and $ \ti\xi_i (x,\ti x)$  
 have been combined
 to form $\x ^M(X)$.
Note that with these definitions the transport derivative takes the form
$\xi^i \pa_i
+ \ti \xi _i \ti \pa ^i  = \x^M \partial_M$.
In this basis the metric $\eta_{MN}$ 
is given by
\be
\eta_{MN}  =\begin{pmatrix}0& I \\ I& 0 \end{pmatrix}  \,.
\ee
We use this metric to raise and lower indices.  We therefore have  
 \be
\partial ^M=   \begin{pmatrix}\, \partial_i\,\\[1.0ex] 
\tilde\partial^i 
 \end{pmatrix}\,  ~~~~
 \x_M=    \begin{pmatrix}\,  \xi^i\,  \\[1.0ex] 
\tilde \xi_i \end{pmatrix}\,.
\ee
We also note that
\be
\Delta = - \eta ^{MN} \partial _M \partial _N  =- \,
\partial^M \partial_M =  - 2 \partial_i \, \tilde \partial^i\,,
 \ee
and therefore  fields $A,B$ restricted to  an arbitrary
isotropic subspace $N$ satisfy 
\be
\label{smallspace1}
\partial^M \partial_M A= \partial^M \partial_M B  
  = 0 \,, ~~~(\partial_M  A) (\partial^M B) =0\,.
\ee
Finally, a short calculation then shows that for any scalar  
operator ${\cal O}$
\be
\label{aux02}
- {1\over 4} \,\Bigl[ \lambda_2 \cdot {\cal O} \lambda_1 - 
\bar\lambda_2 \cdot {\cal O}\,\bar \lambda_1\Bigr] = \,{1\over 2} \,\,
 \x_2^M \,{\cal O}
\, \x_{1M} \,.
\ee

The $2D$-component vectors $\x^M$, $\pa^M$, and $X^M$  all transform
under $O(D,D;\Z)$ by the action of the integer-valued
$2D\times 2D$ matrix $g$
\be
g= \begin{pmatrix}a& b \\ c& d \end{pmatrix}\,,  ~~~g^t \eta g = \eta\,.  
\ee
The C-bracket is covariant under this action of $O(D,D;\Z)$.
If  all dimensions are non-compact so that $\hat M=\R^{2D}$,
 then the continuous group $O(D,D)$ is a  symmetry of the C-bracket while if some of the dimensions are 
 compact,  $M=\R^n\times T^d$,  the symmetry   
 is broken to the subgroup $O(n,n)\times O(d,d;\Z)$
preserving the periodicities of the coordinates.

\medskip
 We now use these relations to rewrite the algebra in an $O(D,D)$ covariant way.
 With the help of (\ref{aux01}) and
(\ref{aux02}) the gauge algebra (\ref{ixltfttasvm}) can be rewritten as
\be
\label{ixltfttaom99}
\begin{split}
~~\lambda^i   _{12}=   
&\,  ~~\x^N_2 \partial_N  \lambda_1^i 
+{1\over 2} \,
\x_2^N \, D^i \, \x_{1N}\,-\, (1\leftrightarrow 2) \,,
\\[1.3ex]
\bar \lambda^i _{12}= & 
~~  \x^N_2 \partial_N \, \bar\lambda_1^i 
- {1\over 2} \,   \x_2^N \, \bar D^i  \x_{1N}\,-\, (1\leftrightarrow 2)~.
\end{split}
\ee
Using (\ref{latoxi}) to define   
parameters $\x_{12}$ and $\tilde \x_{12}$
in terms of $\lambda_{12}$ and $\bar\lambda_{12}$, we readily find that
the above relations imply that
\be
\label{ixltfttvm}
\begin{split}
~~\x^i_{12}=     
&\,   \x^N_2 \partial_N\, \x_1^i 
\,-{1\over 2} \, \,\x_2^N \, \tilde\partial^i \, \x_{1N} \,-\, (1\leftrightarrow 2) \,,
\\[1.3ex]
\tilde \x_{12i}= & \,
    \x^N_2 \partial_N \, \tilde\x_{1i} 
-{1\over 2} \,  \x_2^N \, \partial_i \,  \x_{1N}\,-\, (1\leftrightarrow 2) ~.
\end{split}
\ee
These two relations are summarized by
\be
\label{algfinalform}
\x^M_{12}  =    \x^N_{[2} \partial_N \, \x^M_{1]}
-{1\over 2} \,  \x_{[2}^N \, \partial^M \,  \x_{1]N}\,.
\ee
The algebra is background independent:
the background $E_{ij}$ has dropped out
through the use of appropriate gauge parameters.  The algebra (\ref{algfinalform}) coincides with the one discussed by Siegel in~\cite{Siegel:1993th}.

Equation (\ref{algfinalform})  defines 
the C-bracket (\ref{algfinalform9}) via 
$\x_{12} = - [\x_1, \x_2]_C$.   
For 
 fields $A^M, B^M$
 on 
$\hat M$, we have 
\be
\label{c-bracket5748}
[A, B]_C  =  [A, B]  - {1\over 2} A^P  (\partial ' B_P) + {1\over 2} 
(\partial  ' \hskip-2pt A^P) B_P \,.
\ee
We have introduced the notation $\partial ' $ for the derivative $\partial ^M= \eta ^{MN}\partial _N$ with raised index.
 Here $[A, B]$ is the familiar Lie bracket on the doubled space $\hat M$.
 If the Lie bracket were the only term on the right hand side 
we would have the algebra of diffeomorphisms on $\hat M$.
This is not the case, due to
the extra terms depending on the metric $\eta$ which
lead to new features that will be explored in later sections.

The gauge transformations that leave the fields invariant
have parameters
\be 
 \label{reduciblecova}
 \x^M= \eta ^{MN} \partial _N \chi\,.
 \ee
 As discussed in the introduction, the computation of
 the gauge algebra on fields 
 only determines the algebra up to such terms, so that
 $\x^M_{12}$ is only determined up to the addition of a term
 $\eta ^{MN} \partial _N \chi _{12}$. If this  term   
  is  to be constructed from $\x_1$ and $\x_2$ and involve no further derivatives, $\chi _{12}$ must be of the form
 \be
  \chi _{12}=
\Omega _{PQ} \, \x_{[1}^P \,\x_{2]}^Q \,,  
 \ee
 for some matrix  $\Omega _{PQ}=-\Omega _{QP}$.   
 The general gauge algebra
 is then
 \be
\label{algfinalformxambig}
\x^M_{12}  =    \x^N_{[2} \partial_N \, \x^M_{1]}
-{1\over 2} \, \eta^{MN} \eta_{PQ}\,\x_{[2}^P \, \partial_N \,  \x_{1]}^Q\, 
+ \eta ^{MN} \partial _N \left( \Omega _{PQ}  \x_{[1}^P \x_{2]}^Q \right)
\,.
\ee
In principle $\Omega _{PQ}$ could depend on $x$ and $\ti x$, but
 if the algebra is to have structure constants, as opposed to field-dependent structure functions, 
 it should be a constant matrix.
 Any non-zero choice of  $\Omega _{PQ}$
 will not be invariant under  $O(D,D)$, so  $O(D,D)$
  covariance requires setting $\Omega_{PQ}=0$.

\sectiono{Dilaton, scalars, and vectors} 
\label{dilscavec}

The gauge transformation   (\ref{kssmc})   
 of the dilaton can be written 
covariantly as    
\be
\label{gaugetrans-combo2}
\delta  d  = - {1\over 2}  \partial_N \x^N
+ \x^N \partial_N  \,d\,.
\ee
A short calculation then shows that, for restricted fields, these transformations
satisfy the algebra
 \be
\label{gaugealg-dil}
\bigl[\delta_{\x_1} \,, \delta_{\x_2}\bigr]  d  = - {1\over 2} \, \partial_N {\x}^N_{12}
+ {\x}^N_{12} \,\partial_N  \,d\,,
\ee
with  $\x_{12}$ defined in~(\ref{algfinalform}), so that this is precisely the same gauge algebra found for the transformations of $e_{ij}$.
For $e_{ij}$  we needed to add extra terms to the algebra in order to close the algebra,
but for $d$ the transformations (\ref{gaugetrans-combo2}) close to give precisely the algebra~(\ref{algfinalform}).
We shall take the transformations of $d$ to be    
(\ref{gaugetrans-combo2}) without any higher order corrections.

The transformation  (\ref{gaugetrans-combo2}) can be written as 
\be
\label{gaugetrans-combo3}
\delta  e^{-2d}  =  \partial_N \bigl( \x^N\,e^{-2d}\bigr) \,.
\ee
which is the same as the  transformation of a density $\exp(-2d)$   under a diffeomorphism of $\hat M$
with infinitesimal parameter $\x$.
These transformations, of course, satisfy the
algebra of diffeomorphisms
 \be
\label{diffeo}
\bigl[\delta_{\x_1} \,, \delta_{\x_2}\bigr]    = 
- \d _{{\check\x}_{12}}
\ee
where
\be
{\check\x}_{12}^M\, \equiv \, \x_{[2}^N \, \partial_N  \, \x_{1]}^M 
\ee
 is  
 the Lie bracket 
   $[\x_2,\x_1]$.
 In fact the transformation of $d$ is consistent with both
 the diffeomorphism algebra and the C-algebra~(\ref{algfinalform}) 
 because  for restricted fields use of 
  (\ref{smallspace1})  leads to
\be
\label{divinss}
\partial_M \x^M_{12}  = \partial_M {\check\x}_{12}^M ~~~\hbox{and} ~~~ \x^M_{12} \partial_M
= {\check\x}_{12}^M  \partial_M \,.
\ee
Then the algebra of transformations on $d$ can 
 also be written as
 \be
\label{gaugealg-dil}
\bigl[\delta_{\x_1} \,, \delta_{\x_2}\bigr]  d  = - {1\over 2} \,
 \partial_N\, {\check\x}^N_{12}
+ {\check\x}^N_{12} \,\partial_N  \,d\,.
\ee

It is clear from the above discussion that one can define scalars as well as densities
in this theory.  A scalar $R$ is required to  transform as
\be
\delta_\x R =  \x^N \partial_N  \, R \,.
\ee
The resulting gauge algebra (\ref{diffeo}) is that of diffeomorphisms, and this
is consistent  
with~(\ref{algfinalform})
since $  {\check\x}^N_{12} \,\partial_N R= {\x}^N_{12} \,\partial_NR$ for restricted fields.

  With a scalar $R (e,d)$ built
using the fields $e_{ij}$ and $d$, a gauge invariant action (for restricted fields) could be
constructed as
\be
S = \int dx d\ti x  \, e^{-2d}  \, R (e,d) \,.  
\ee
Since the fields are restricted to some $D$-dimensional null torus $N$, the
integral in the action could be restricted to $N$.
To leading order in the fields, a suitable candidate with two
derivatives is 
\be
R(e,d) = 4 D^2 d +  D^i \bar D^j e_{ij} + \hbox{quadratic in fields}\,.
\ee
An all orders construction of this scalar would be very useful.
In the work~\cite{Siegel:1993th} of Siegel, a scalar is constructed and presumably should
agree with $R(e,d)$, once suitable gauge conditions are imposed
to eliminate the extra gauge degrees of freedom in that 
formulation.

We conclude with a brief discussion of vectors.
A vector field $V^M$
on $\hat M$ transforms under an infinitesimal diffeomorphism on  $\hat M$ with the Lie bracket $\delta_\x V = [\x , V ]$.  The   parameter $\x^M$ is also a vector field on  $\hat M$ and the Jacobi identities ensure this is a representation of the diffeomorphism algebra. 
It is  straightforward 
 to see that such a vector
 \be
  V^M \equiv    \begin{pmatrix}\, \ti v_i(x,\tilde x)\, \\[1.0ex] 
v^i (x, \tilde x) \end{pmatrix}\,,
 \ee
  restricted to
$M$ results in a $v^i(x)$ that  is a vector field on $M$ but gives a 
$\ti v_i (x)$   
that does not transform as a 1-form, but rather as a scalar   
 under diffeomorphisms of $M$.  
 This means that
 despite the suggestive notation, 
 our gauge parameters $\x^M$ should not be thought of as 
 conventional vector fields on $\hat M$, as their restriction gives a vector field $\xi(x)$ and a 1-form $\ti \xi (x)$, not a vector and scalar.
In fact, the reducibility of the symmetry means that, for the restriction to $M$,
$\ti \xi $ and $\ti \xi +d \a$ generate the same transformations.
Then $\ti \xi $ and $\ti \xi +d \a$ 
   can be regarded as equivalent, so that  $\ti \xi $ is more properly thought of as a 1-form connection on $M$.

 It is natural to attempt 
  a generalisation of vectors using the C-bracket.  A C-vector $V$ on $\hat M$
 would then transform as
 \be
 \label{c-vector}
 \delta_\x  V = [\x \,, V ]_{{}_C} \,. 
 \ee
The algebra of these transformations would be exactly~(\ref{algfinalform})
if the Jacobi identity held for the C-bracket.  Since it does not, and the
Jacobiator is of the form  $\partial^M \chi$, the consistency of the algebra  requires that vectors
are only defined up to the equivalence  $V^M \sim V^M + \partial^M \chi$, so that 
$V_M$ are the components of a connection one-form on $\hat M$.
In particular, the gauge parameters $\x^M(X)$ are  such  C-vectors on $\hat M$, as parameters
$\x^M$ and $ \x^M + \partial^M \chi$ define the same transformations and so are equivalent.

\sectiono{Restricted fields}\label{restridual}

\medskip   
The original space $M= T^D$  
 has coordinates $x^i$ and the dual space $\ti M= T^D$ has coordinates 
 $\ti x_i$.
Together they form the doubled space $\hat M= M\times \ti M
= T^{2D}$.  
The parameters $\xi (x,\ti x)$ and $\ti\xi (x,\ti x)$
combine to form $\x ^M(X)$.
The metric $\eta$   
 of signature  $(D,D)$ is
$ds^2    
= 2dx^i d\ti x_i\,.$   
 Fields $A$  
 restricted to satisfy
\be
\label{Mrestrict}
{\partial \over \partial \ti x_i} A=0\,,
\ee
will be referred to as geometric, as they are fields on the spacetime $M$.

\medskip
We   now
consider the restriction to 
a general isotropic subspace, namely,  
a $D$-dimensional  torus submanifold $N\subset \hat M$
that is null with respect to $\eta$.
Then $\hat M= N\times \ti N$, where $\ti N$ is another $D$-dimensional null torus.
We introduce periodic coordinates $y^i$ on $N$ and $\ti y_i $ on $\ti N$ ($i,j=1,...,D$)
so that the metric takes the form
$ds^2  
=2dy^i d\ti y_i $.
Following \cite{Hull:2004in,Hull:2006va}, 
we can describe the
choice of $N$ with  constant projectors
$\Pi$ and  $\ti \Pi$:
\be
y^i = \Pi ^i {}_M X^M, \qquad \ti y_i = \ti \Pi_{iM} X^M \,.
\ee
As these preserve the metric and respect the periodicities of all coordinates, 
they define an $ O(D,D;\Z)$ transformation $\Phi$:
\be
\label{oddrot}
   \begin{pmatrix} \,\ti  y \,\\[0.5ex] 
y \end{pmatrix}
= \Phi
   \begin{pmatrix} \,\ti x \,\\[0.5ex] 
x \end{pmatrix}\,, ~~~~ 
\Phi^{  I} {}_J= \begin{pmatrix}
\, \ti \P_{iJ}\, 
  \\[0.8ex]  
   \P ^i{}_J
  \end{pmatrix}\,.
 \end{equation}
The restriction of fields now takes the form
 \be
  \label{Nrestrict}
{\partial \over \partial \ti y_i} A=0 \,.
\ee
Restricted parameters $\x^M(y)$ are fields on $N$
and the projectors can 
be used to decompose
$\x^M(y)$ into a vector field $\z^i(y)$ on $N$ and a one-form field $\ti \z_i(y)$ on $N$:
\be
\label{xisz}
\z^i = \Pi ^i {}_M \x^M, \qquad
\ti \z_i= \ti \Pi_{iM}\x^M\,. 
\ee

The restriction  (\ref{Nrestrict}) 
to a subspace $N$  clearly breaks the $O(D,D;\Z)$ symmetry.
For the restriction to the spacetime $M$, the constraint 
${\partial \over \partial \ti x_i} A=0 $ (see (\ref{Mrestrict}))
is  preserved by the $GL(D;\Z)$ subgroup of $O(D,D;\Z)$
\be
\label{adsfa}
g= \begin{pmatrix}a& 0 \\ 0& \ti a \end{pmatrix}\,,
\ee
where $a\in GL(D;\Z)$ and $\ti a \equiv  (a^t)^{-1}$.  Indeed, these 
transformations simply rotate the $\partial_i$ and $\ti \partial^i$ derivatives
among themselves.   The constraint (\ref{Mrestrict}) is also preserved
by the $B$-transformations   
\be
\label{brtrans}
g= \begin{pmatrix}1& \theta \\ 0& 1 \end{pmatrix}\,.   
\ee
where $\theta $ is a constant integer-valued antisymmetric  matrix. This transformation
 acts on the derivatives~as
\be
\pa _i \to \pa _i + \theta_{ij}\,\ti \pa ^j, \qquad  \ti \pa ^i \to  \ti \pa ^i,
\ee
making it clear that the constraint is 
unchanged.

The above $GL(D;\Z)$ and $B$-transformations form the
 \lq geometric' subgroup~$\G$ of $O(D,D;\Z)$~\cite{Hull:2004in}.
The restriction (\ref{Nrestrict})  to $N$  is preserved by a group  $\G_N$
conjugate to the geometric subgroup $\G$:
\be
\G_N= \{ \Phi g \Phi ^{-1} : g\in \G\}\,.
\ee

For  $M={\cal M}_n\times T^d$, the product of $n$-dimensional Minkowski space  ${\cal M}_n$ with coordinates $x^\mu$ and a $d$-torus  with coordinates $x^a$, we introduce a dual space
$\ti M=  \ti  {\cal M}_n\times T^{d}$
with a dual 
$n$-dimensional Minkowski space  $\ti {\cal M}_n$ with coordinates $\ti x_\mu$ and a dual $d$-torus  with coordinates $\ti x_a$.
 The doubled space is
$\hat M=M\times \ti M= {\cal M}_n\times \ti  {\cal M}_n\times T^{2d}
$. 
Absence of winding in the non-compact directions requires that all fields and parameters are independent of $\ti x_\mu$.
Restricted fields must then 
 be fields on a space $N={\cal M}_n\times N_d  
 \subset \hat M$,  
where $N_d$ is a $d$-dimensional
null torus subspace of the double torus $T^{2d}$.
Then there is a null  
torus  $\ti N_d$ so that $T^{2d}=N_d\times \ti N_d$. If the coordinates of $N_d$ are $y^a$ and those of $\ti N_d$ are $\ti y_a$, then
 fields restricted to $N$ are independent of
 $\ti y_i= (\ti x_\mu, \ti y_a)$.  
 
\sectiono{Reducibility and the Jacobiator }\label{Jacalg}

In this section, we discuss further the issues concerning a gauge algebra with  a non-vanishing Jacobiator and
the ambiguities in the gauge algebra. 
Consider then a (possibly infinite dimensional) closed algebra
\be
\label{alg}
[T_A, T_B]=f_{AB}{}^CT_C\,,
\ee
with constant $f_{AB}{}^C$. If the $T_A$ are to be realised as a set of classical infinitesimal symmetry transformations on a set of fields, with $[T_A, T_B]$ the commutator of two such transformations, then it is necessary that
\be
\label{jaco}
[[T_A,T_B],T_C] + {\rm cyclic~ permutations}
\ee
vanishes when acting on fields. This will be the case if the structure constants satisfy the Jacobi identity
so that we have a Lie
algebra. However, suppose that (\ref{jaco}) is not zero, but takes values in a closed sub-algebra
with basis $Z_\alpha$.  Let a basis for the remaining generators be $t_a$, so that 
$T_A=\{ t_a, Z_\a \}$.
Then
the structure constants satisfy
\be
\label{violated}
f_{[AB}{}^Df_{C]D}{}^a=0, \qquad f_{[AB}{}^Df_{C]D}{}^\a=-
2 g_{ABC}{}^\alpha  \,,   
\ee
for some constants $g_{ABC}{}^\alpha $, 
so that the generators   satisfy
\be
[[T_A,T_B],T_C] + {\rm cyclic~ permutations}=
g_{ABC}{}^\alpha Z_\alpha  \,.    
\ee
Then the algebra can be realised as 
classical infinitesimal symmetry transformations on a set of fields provided that the subalgebra of the $Z$'s leaves all the fields invariant, so the symmetry  is reducible. If each $Z_\a$ gives zero when acting on every field, then 
 (\ref{jaco}) will give zero acting on fields, as required.\footnote{In the  Batalin-Vilkovisky master action, $g_{ABC}{}^\alpha $ appears in the quartic  coupling of three ghost fields to the anti-field of a 2nd generation ghost.}
 
 A general infinitesimal  transformation will be a linear combination  $\Sigma^AT_A$ of the $T_A$
 for some $\Sigma^A$. As usual, the algebra (\ref{alg}) defines a bracket of the parameters $\Sigma^A$
 \be
 \label{braccc}
[\Sigma_1 \,, \Sigma_2]^D= 
f_{AB}{}^D \Sigma_1 ^A\Sigma_2^B \,, 
 \ee
 with Jacobiator
 \be
\begin{split}  {J(\Sigma_1, \Sigma_2, \Sigma_3)} ^E=  
\frac 1 2
\Sigma_1 ^A\Sigma_2^B\Sigma_3^C
 \, f_{[AB}{}^Df_{C]D}{}^E   \,,   
 \end{split}
\ee
so that 
  \be   
\begin{split}  
{J(\Sigma_1, \Sigma_2, \Sigma_3)} ^a   
=  0\,, \qquad \quad 
 {J(\Sigma_1, \Sigma_2, \Sigma_3)} ^\a 
 =  -\Sigma_1 ^A\Sigma_2^B\Sigma_3^C
 g_{ABC}{}^\alpha 
  \,.
\end{split}
\ee
 
 Our symmetry algebra is of precisely this type. The parameters are $ \Sigma^M(x,\ti x)$ so that the corresponding generators are $T_M(x,\ti x)$, with the classical transformation  $\d_\x$ with  infinitesimal  parameter $ \Sigma^M(x,\ti x)$ written as $\int dx d\ti x \, \Sigma^M(x,\ti x)T_M(x,\ti x)$. It is convenient to write this as  $\Sigma^AT_A$ with $A$   a composite index representing the discrete index $M$ and the continuous variables $x,\ti x$ and summation over $A$ representing summation over  $M$ and integration over   $x,\ti x$.
 The 
 structure constants $f_{AB}{}^C$ are then defined by the  C-bracket through (\ref{braccc}) where $[\Sigma_1 \,, \Sigma_2]$ is minus the C-bracket.
 There is a subalgebra of generators $Z_\alpha$ generating transformations that leave the fields invariant, consisting of   transformations 
$\Sigma^AT_A$ with $\x$ of the form 
 (\ref{reduciblecov}). The structure constants for the Courant bracket and the C-bracket satisfy relations of the form 
(\ref{violated}) so that the Jacobi identitites are violated by terms in the $Z$-algebra. As these do not act on the fields, the algebra can be consistently realised on fields.
It is tempting to try to set the $Z$-generators to zero in some way, but there does not appear to be a 
local covariant 
way of doing this.
Reducibility  is intimately related to the failure of the Jacobi identities.

The algebra (\ref{alg}) can be written as
\be
\label{zalg}
[T_A, T_B]=f_{AB}{}^ct_c 
+f_{AB}{}^\alpha Z_\alpha  \,.   
\ee
The calculation of the algebra of gauge transformations $[\d _{\x_1}, \d _{\x_2}]$ acting on fields
only determines the structure constants $f_{AB}{}^c$ but leaves the $f_{AB}{}^\alpha$ completely undetermined as $Z_\a$ does not act on fields.
In our case, demanding $O(D,D)$ covariance fixes the
$f_{AB}{}^\alpha$ completely and gives the algebra of the C-bracket.
We shall see, however, that other non-covariant choices are possible, including one which gives an algebra that does satisfy the Jacobi identity.

\sectiono{The Courant bracket}\label{courantbrackets}

The Courant bracket is 
 defined on smooth
sections of $T\oplus T^*$, where $T$ is the tangent bundle of a manifold and $T^*$ the cotangent bundle. Such a section is  the formal sum $A+\alpha$ of a vector field $A$ and a 1-form field $\a$.
For two such sections $A+\alpha$ and $B+\beta$,  
 the Courant bracket is the skew-symmetric bracket  given by
\begin{equation}
\label{cbracket}
[A+\alpha,B+\beta]_{{}_{\rm{Cour}}}  = [A,B]+ \LL_A\beta- \LL_B\alpha\,-\frac{1}{2}d(i_A\beta-i_B\alpha).
\end{equation}
Note that for two such sections, there is also a natural inner product 
\be
\label{innprod}
\langle {A+\alpha,B+\beta}\rangle=
i_A\beta+i_B\alpha \,, \ee
which is of signature $(D,D)$ on a space of dimension $D$. This is the flat metric $\eta$.

The Courant bracket is not a Lie bracket since it fails to
satisfy the Jacobi identity. 
The Jacobiator (see (\ref{jacobiator-def})) of sections $P, Q, R$ of
$T\oplus T^*$  
 is given by~\cite{LWX, Gualtieri}:
\begin{equation}
\label{jacob-courant}
J(P,Q,R) 
= d\, N(P,Q,R) ,
\end{equation}
 where $N$ is the \lq Nijenhuis operator'  given by
\begin{equation}
N(P,Q,R)=
 \frac{1}{6}
\,\Bigl(\langle {[P,Q]_{{}_{\rm{Cour}}},R}\rangle 
+\langle{[Q,R]_{{}_{\rm{Cour}}}\,, P}\rangle   
+\langle {[R,P]_{{}_{\rm{Cour}}},Q}\rangle\Bigr)\,.   
\end{equation}

While the gauge algebra for diffeomorphisms is given by the Lie bracket, that for diffeomorphisms plus $b$-field gauge transformations has a subtlety that is suggestive of the Courant bracket.  
Under an infinitesimal diffeomorphism with 
a vector field parameter  
$\xi$ and a 
$B$-field gauge transformation with 
a one-form parameter $\ti \xi$ we have
\be
\delta g=  {\cal L}_\xi g\,, ~~~\hbox{and} ~~~
\delta b = {\cal L}_\xi b + d\ti  \xi\,,
\ee
for a metric $g$ and a two-form field~$b$. 
The symmetry is reducible: replacing $\ti \xi \to \ti \xi + d \s$ leaves the transformations unchanged and constitutes a ``symmetry for a symmetry".
The computation of the
algebra quickly gives the first three terms of the right-hand side 
of~(\ref{cbracket}).  
The last term, $d(\ldots)$, can be added with
{\em arbitrary} coefficient, as it represents an ambiguity:  the gauge transformations are unchanged
when the 
$B$-field gauge parameter changes by an exact term. 
This is the ambiguity discussed in the introduction and \S\ref{Jacalg}.
In (\ref{zalg})  the structure constants 
  $f_{AB}{}^\alpha$ are completely undetermined as $Z_\a$ does not act on the fields.
  Here $Z_\a$ generate transformations with $ \ti \xi = d \s$.
Hence the gauge 
algebra acting on fields gives the bracket
\begin{equation}
\label{bbracket}
[A+\alpha,B+\beta]_\k  = [A,B]+ \LL_A\beta- \LL_B\alpha\,-\frac{1}{2}\,\k\, d(i_A\beta-i_B\alpha).
\end{equation}
with arbitrary coefficient $\k$.
Taking $\k=0$ gives an  
algebra that satisfies the Jacobi identity.
Any $\k\ne 0$ gives a non-zero Jacobiator equal to an exact one-form,
so that  the resulting gauge parameter does not act on any fields.   
Taking $\k=1$ gives the Courant bracket.
 
\medskip
We can now show that the C-bracket reduces to the
Courant bracket when the parameters are required to 
be independent of $\ti x$.  To do this we write out the terms
in (\ref{algfinalform9})  
\be
\label{Cbrack}
\begin{split}
\bigl([\x_1 , \x_2]_{{}_C}\bigr)^M 
 &=  \x_{1}^N \partial_N \x_{2}^M -  \x_{2}^N \partial_N \x_{1}^M  
~ - {1\over 2} \, \x_{1}^N\,\partial^M  
\x_{2N}  + {1\over 2} \, \x_{2}^N\,\partial^M  
\x_{1N} \,.  
\end{split}
\ee
To find an explicit formula in terms of $\xi, \ti \xi$ and $\pa, \ti \pa$ 
we use the definitions at the beginning of \S\ref{exploring}.
Since 
the parameters are restricted to be independent of $\ti x$, 
all terms involving $\ti \pa ^i$ in (\ref{Cbrack}) vanish and $\xi^i(x)$ is a vector field on $M$ and $\ti \xi _i(x)$ is a 1-form field.
Then 
\be
\bigl( [\x _1, \x _2]_{{}_C}\bigr)^M  =
\begin{pmatrix} ([\x_1, \x _2]_{{}_C})_i   \\[0.8ex] ([\x_1, \x _2]_{{}_C})^i 
\end{pmatrix}  \,,
\ee
where
\be 
\bigl([\x_1 , \x_2]_{{}_C}\bigr)^i =  \xi_1^j \partial_j \xi^i_2 -   \xi_2^j \partial_j 
 \xi^i_1 =   (\cL_{\xi_1}\xi _2)^i  =  \bigl( [\xi_1 , \xi_2 ]\bigr)^i\,,
 \ee
 is the Lie bracket of two vector fields
 while 
 \be 
 \begin{split}
 \bigl( [\x_1, \x _2]_{{}_C}\hskip-1pt\bigr)_i
  &=
  \xi _1^j \pa_j \ti \xi_{2i}\,-{1\over 2}(\xi_1^j \pa_i \ti \xi _{2j} -
 \ti \xi _{2j} \pa_i \xi^j_1 ) \quad - ( 1 \leftrightarrow 2)
 \\
 &= \xi _1^j \pa_j \ti \xi_{2i}\,
+ (\pa_i \xi^j_1) \ti \xi _{2j} \,   -{1\over 2}\pa_i(\xi_1^j \, \ti \xi _{2j} ) \quad  - ( 1 \leftrightarrow 2)
\\
&=
\Bigl(\,\, \cL_{\xi_1} \ti \xi_2   -\half
\, d ( i_{\xi_1}\ti \xi _{2} )~\Bigl)_i   ~~ - ( 1 \leftrightarrow 2)\,.
\end{split}
  \ee
Rewriting the above results 
in  terms of the formal sum $ \xi +\ti \xi$, we find that 
the C-bracket for parameters independent of $\ti x$ is precisely the Courant bracket 
$[\xi_1+\ti \xi _1, \xi_2+\ti \xi _2]_{{}_{\rm{Cour}}}$.
 A very similar calculation shows that for parameters restricted to $N$ by (\ref{Nrestrict}), the C-bracket $[\x_1,\x_2]_C$ is precisely the Courant bracket  
$[\z_1+\ti \z_1,\z_2+\ti \z _2]_{{}_{\rm{Cour}}}$ on $N$ 
where the parameters have been decomposed into vectors $\z$ and one-forms $\ti \z$ on $N$ as in (\ref{xisz}).
This is of course as was to be expected, as the restriction to $N$ is obtained from the restriction to $M$ by the $O(D,D;\Z)$ transformation (\ref{oddrot}) which 
is a symmetry of the C-bracket, so that the $M$ restriction and the $N$ restriction are isomorphic.

As discussed at the end of \S\ref{exploring},   
calculating the gauge algebra on the fields gives
the gauge algebra (\ref{algfinalformxambig}) with any choice of 2-form $\Omega _{PQ}$.
If we choose
\be
 \Omega = \gamma \begin{pmatrix} \phantom{-}0& 1 \\ -1& 0 
 \end{pmatrix} \,,
\ee
then we obtain
 \be 
 \begin{split}
  [\x_1, \x _2]_{{}_C i}
 &=
\Bigl(\,\, \cL_{\xi_1} \ti \xi_2   -\half \k
\, d ( i_{\xi_1}\ti \xi _{2} )~\Bigl)_i  ~  - ( 1 \leftrightarrow 2)\,,
~~\hbox{with}~~  \k = 1 +4\gamma \,.   
\end{split}
  \ee
The Courant bracket has been replaced by the 
$\k$-bracket~(\ref{bbracket}). 
 This is of course to be expected, as the two systems are related by field redefinitions, as we showed in \S\ref{thegaugetransformations}.

 These brackets also arise from current algebras.
 In the canonical treatment of the string in flat space with coordinates $x^i$, the canonical variables
 are loops $x^i(\s)$
 with conjugate momenta $p_i(\s)$. 
 Here $\s$ is a periodic coordinate on the string.
The currents
\be
J_{\xi +\ti \xi }=\xi ^ip_i +\ti \xi _i {\frac {dx^i} {d \s}} \, ,
\ee
satisfy a canonical current algebra
\be
[J_{\xi +\ti \xi }, J_{\z +\ti \z }
]=J_{\chi +\ti \chi }+\dots
\ee
where
\be 
{\chi +\ti \chi }
=[{\xi +\ti \xi }, {\z +\ti \z }
]
\ee
defines a bracket.
In  \cite{Alekseev:2004np}
this calculation was done for the case in which
$\xi$ and $\ti \xi$ depend on $x(\s)$ and it was found that the resulting bracket could be the Courant bracket. In fact there is an ambiguity in the calculation and the general result is in fact the $\k$-bracket given above.
  Siegel had done essentially the same calculation earlier 
 \cite{Siegel:1993th} in the more general context in which $\xi$ and $\ti \xi$ 
 depend on both $x$ and $\ti x$ but are restricted to take values on some $N$. In this case, he 
  used a duality covariant formalism and
 found precisely the C-bracket.

\sectiono{The C-Bracket}\label{cbracksec}

We have seen that the C-bracket of fields restricted to the null $D$-dimensional torus $N$ in $\hat M= M\times \tilde M$  is in fact the
 Courant bracket on $(T\oplus T^*)N$.  
In this section we 
write the C-bracket in terms of 
derivatives with respect to coordinates $x^i$ of $M$ and coordinates
$\tilde x_i$ of $\tilde M$. 
We decompose  each  
$\x^M$ on $\hat M$ into $\xi ^i(x,\ti x)$ and   $\ti \xi_i(x,\ti x)$
as in (\ref{vecx}). 
We find a symmetric
structure in which $\xi$  and $\ti \xi$ are treated similarly.
 If $\x$ is restricted  to
 $M$, then $\xi ^i(x)$ and  $\ti \xi_i(x)$
are vectors  and one-forms  on $M$.

The asymmetry in the treatment of vectors and one-forms in the
Courant bracket led the authors of~\cite{LWX}  to introduce a Courant-like bracket treating 
them symmetrically.
With sections $A, B$ of a bundle $L$ and 
sections $\a, \b$ of a dual bundle $L^*$, where
$(L, L^*)$ form a so-called Lie bi-algebroid,  the 
 bracket takes the form:
\be
\label{bi-alg}
\begin{split}
[A+\alpha,B+\beta]&~=~~ [A,B] +
 \LL_\alpha B -   
 \LL_\beta A
+\tfrac{1}{2}\ti d (i_A\beta-i_B\alpha)\\[1.0ex]
&~~~+[\alpha,\,\beta] \,+ \LL_A\beta- \LL_B\alpha\, -
\tfrac{1}{2}d(i_A\beta-i_B\alpha).
\end{split} 
\ee
This bracket is the key element in making $L\oplus L^*$ into a ``Courant
algebroid".
We show  that the C-bracket, with natural definitions associated with
the space $M\times \tilde M$,    takes precisely  the
form~(\ref{bi-alg}).   This is true even if the fields are not restricted!
However, it is only when the fields are restricted to $N$ that this becomes an example of the general setup of~\cite{LWX}, giving a Courant algebroid over $N$.\footnote{ $(TN,T^*N) $  form a Lie bi-algebroid over $N$, and their sum is a Courant algebroid  $(T\oplus T^*)N$ over $N$ with bracket (\ref{bi-alg}), which is the Courant bracket on $TN \oplus T^* N$.}
We conclude with a computation of  the Jacobiator of the C-bracket.
We assume restricted fields, but  make no explicit reference to the choice of $N$. We show that the Jacobiator is
a trivial gauge parameter, as required from the discussion given in the introduction and~\S\ref{Jacalg}.

\bigskip

We begin by considering the C-bracket, 
which takes the form~(\ref{Cbrack}):
\be
\label{brack}  
\begin{split}
\bigl( [\x_1 , \x_2]_{{}_C}\bigr)^M  &= \bigl( [\x_1 , \x_2]\bigr)^M  
~ - {1\over 2} \, \x_{[1}^N\,\partial^M  
\x_{2]N}  \\[1.0ex]
 &=  \x_{1}^N \partial_N \x_{2}^M -  \x_{2}^N \partial_N \x_{1}^M  
~ - {1\over 2} \, \x_{1}^N\,\partial^M  
\x_{2N}  + {1\over 2} \, \x_{2}^N\,\partial^M  
\x_{1N} \,,  
\end{split}
\ee
where $[\x_1 , \x_2]$ is the Lie bracket for the doubled space $\hat M$.
Each $\x$  
is decomposed into
a $\xi ^i$ and a $\ti \xi_i$ as in (\ref{vecx}). 
For notational convenience we write
$\x = \xi  + \tilde \xi$ formally adding together 
$\xi$  and
$\tilde \xi $. 
The bracket
can then be evaluated as
\be
\label{mdffj}
[\xi_1 + \tilde\xi_1\,, \xi_2 +  \tilde \xi_2]_{{}_C} =  [\xi_1, \xi_2]_{{}_C} + [\xi_1, \tilde\xi_2]_{{}_C} + [\tilde \xi_1\,, \xi_2]_{{}_C}
+ [\tilde\xi_1, \tilde \xi_2]_{{}_C}\,.
\ee
  For $[\xi _1, \xi _2]_{{}_C}$ and $[\ti \xi _1, \ti \xi _2]_{{}_C}$, the second term  in the algebra (\ref{brack}) vanishes because this term necessarily couples
a $\xi $ to a $\ti \xi$  as  the metric $\eta$ is off-diagonal.
Therefore, for both these computations the C-bracket  
 reduces to the
Lie bracket on the doubled space.  Let us first consider $[\xi _1, \xi _2]_{{}_C}$.  We have
\be
\bigl( [\xi _1, \xi _2]_{{}_C}\bigr)^M  = \bigl([\xi _1, \xi _2]\bigr)^M= 
\begin{pmatrix} ([\xi _1, \xi _2]_{{}_C})_i\\[1.1ex] ([\xi _1, \xi _2]_{{}_C})^i 
\end{pmatrix}  =
 \xi_1^j \partial_j   
\begin{pmatrix} 0\\ \xi^i_2 \end{pmatrix} 
- \xi_2^j \partial_j 
\begin{pmatrix} 0\\ \xi^i_1 \end{pmatrix} \,.
\ee
We thus conclude that
\be
\label{twotildes9}
\begin{split}
\bigl( [\xi _1, \xi _2]_{{}_C}\bigr)_i &=  ~~0\,, \\[0.5ex]
\bigl( [\,\xi _1, \xi _2\,]_{{}_C}\bigr)^i &=  \xi_1^j \partial_j \xi^i_2 -   \xi_2^j \partial_j 
 \xi^i_1
 \,.
\end{split}
\ee
With the Lie derivative
\be
  (\cL_{\xi_1}\xi _2)^i  =\xi_1^j \partial_j \xi^i_2 -   \xi_2^j \partial_j 
 \xi^i_1 \,,   
 \ee
and the bracket 
$ ([\xi_1 , \xi_2 ])^i= (\cL_{\xi_1}\xi _2)^i$, we have 
\be
\label{eq01}
[\xi _1,  \xi _2]_{{}_C }=  \cL_{\xi_1} \xi _2 = [\xi_1 , \xi_2]\,.
\ee
For fixed $\ti x$, this would be the usual Lie derivative and Lie bracket on $M$.

Let us now consider $[\ti \xi _1, \ti \xi _2]_{{}_C}$.  This time we get
\be
\bigl([\ti \xi _1, \ti \xi _2]_{{}_C}\bigr)^M  = \bigl([\ti\xi _1, \ti\xi _2]\bigr)^M= 
\begin{pmatrix} ([\ti\xi _1, \ti \xi _2]_{{}_C})_i\\[1.1ex] 
([\ti\xi _1, \ti \xi _2]_{{}_C})^i 
\end{pmatrix}  =
 \ti\xi_{1j} \ti\partial^j   
\begin{pmatrix} \ti\xi_{2i} \\ 0 \end{pmatrix} 
- \ti\xi_{2j}\ti \partial^j 
\begin{pmatrix} \ti\xi_{1i}\\0  \end{pmatrix} \,.
\ee
giving
\be
\label{twotildes}
\begin{split}
\bigl([\ti\xi _1, \ti \xi _2]_{{}_C}\bigr)_i &=  \ti\xi_{1j} \ti\partial^j \ti\xi_{2i}
-  \ti\xi_{2j}\ti \partial^j  \ti\xi_{1i} \,,\\[0.8ex]
\bigl([\ti\xi _1, \ti \xi _2]_{{}_C}\bigr)^i & = ~0\,. 
\end{split}
\ee
For any
 $\alpha_i $ and $\beta _i$ we define the Lie derivative
along $\alpha$ of $\beta$ as
\be
\label{deftilie}
(
\cL_{\alpha} \beta )_i \equiv  \alpha_j \ti\partial^j\, \beta_i  -  (\ti\partial^j \alpha_i) \beta_j =   \bigl([\alpha\,, \beta ]\bigr)_i \,.
\ee
The Lie-bracket $[\alpha\,, \beta ]$ has a lower index and is
computed using only $\tilde x$-derivatives. 
Then we have 
\be
\label{eq02}
[\ti\xi _1, \ti \xi _2]_{{}_C }= 
\cL_{\ti \xi_1}\ti \xi _2 = [\ti\xi_1 , \ti \xi_2]\,.
\ee

\medskip
The mixed terms bring 
new features.
Let us compute $[\xi_1, \ti \xi_2 ]_{{}_C}$.  This time we get contributions
from all terms in the C-bracket:
\be
\bigl([\xi _1, \ti \xi _2]_{{}_C}\bigr)^M  = 
\begin{pmatrix} ([\xi _1, \ti \xi _2]_{{}_C})_i\\[1.1ex] 
([\xi _1, \ti \xi _2]_{{}_C})^i 
\end{pmatrix}  =
 \xi_1^j \partial_j   
\begin{pmatrix} \ti\xi_{2i} \\ 0 \end{pmatrix} 
- \ti\xi_{2j}\ti \partial^j 
\begin{pmatrix} 0\\ \xi_1^i  \end{pmatrix} 
-{1\over 2} \, \xi_1^j  \begin{pmatrix} \partial_i \\[0.5ex] \ti\partial^i 
\end{pmatrix} \ti \xi_{2j}
+{1\over 2} \, \ti\xi_{2j}  \begin{pmatrix} \partial_i \\[0.5ex] \ti\partial^i 
\end{pmatrix}  \xi_1^j
\,.
\ee
We thus get
\be
\label{mixedterms}
\begin{split}
\bigl([\xi_1, \ti \xi_2]_{{}_C}\bigr)_i &=~\,
\xi _1^j \pa_j \ti \xi_{2i}\,-{1\over 2}(\xi_1^j \pa_i \ti \xi _{2j} -
 \ti \xi _{2j} \pa_i \xi^j_1 )~=~~ \xi _1^j \pa_j \ti \xi_{2i}\,
+ (\pa_i \xi^j_1) \ti \xi _{2j} \,   -{1\over 2}\pa_i(\xi_1^j \, \ti \xi _{2j} )\,,
 \\[0.5ex]
\bigl([\,\xi_1, \ti \xi_2\,]_{{}_C}\bigr)^i &=- \ti \xi _{2j} \ti \pa ^j \xi_1^i
-{1\over 2}(\xi_1^j \ti \pa^i \ti \xi _{2j} -
 \ti \xi _{2j}\ti \pa^ i \xi_1^j ) 
 =- \ti \xi _{2j} \ti \pa ^j \xi_1^i  -(\ti \pa^i \ti \xi _{2j})\xi_1^j  + {1\over 2}\ti \pa^i(\xi_1^j\,  \ti \xi _{2j} )\,.
 \end{split}
\ee
A few natural definitions help rewrite this more clearly.  Given    
$A^i,\alpha_i$ we define
\be
\label{mrlie}
\begin{split}
(\cL_A \alpha)_i &\equiv A^j \partial_j \,\alpha_i + (\partial_i A^j) \alpha_j
\,, \\[0.8ex]
(
\cL_\alpha A)^i &\equiv   \alpha_j \ti\partial^j A^i  + (\ti\partial^i \alpha_j) A^j\,.
\end{split}
\ee
We also define exterior derivatives $d$ and $\ti d$.  Acting
on a function 
 $S$ they give  
\be
\begin{split}
(dS)_i &= \partial_i S \,, ~~~(dS)^i = 0\,,  \\[0.4ex]
(\ti d S)^i  &=  \ti\partial^i S \,, ~~~(\ti d S)_i = 0 \,. 
\end{split}
\ee
Finally, we   define contractions and dual contractions,
\be
\alpha_i A^i = i_A \alpha =  \ti \i_\alpha A \,. 
\ee
One can verify that  on any $A^i$ (as in the second equation in (\ref{mrlie})),
we have
\be
\ti \cL_\alpha  =  \ti d \, \ti \i_\alpha  + \ti \i_\alpha  \ti d \,.
\ee
This is completely analogous to the formula that gives the action of 
standard Lie derivatives on forms.  We can now return to 
(\ref{mixedterms}) and write
\be
\label{mixedterms99}   
\begin{split}
\bigl([\xi_1, \ti \xi_2]_{{}_C}\bigr)_i &=~\Bigl(\,\, \cL_{\xi_1} \ti \xi_2   -\half
\, d ( i_{\xi_1}\ti \xi _{2} )~\Bigl)_i\,,
 \\[0.9ex]
\bigl([\,\xi_1, \ti \xi_2\,]_{{}_C}\bigr)^i &
 = \Bigl( - 
 \cL_{\ti\xi _2}  \xi_1  + \half\ti d (\ti \i_{\ti\xi_2} \xi_1 )
 \,\Bigr)^i\,.
 \end{split}
\ee
The two expressions above are summarized 
by
\be
\label{fldke}
[\xi_1, \ti \xi_2]_{{}_C} = - 
\cL_{\ti\xi _2}  \xi_1  + \half \ti d (\ti \i_{\ti\xi_2} \xi_1 )~+~ \cL_{\xi_1} \ti \xi_2   -\half
\, d ( i_{\xi_1}\ti \xi _{2} )\,.
\ee
Using the above result, together with (\ref{eq01}) and (\ref{eq02}) we readily evaluate (\ref{mdffj}).  The result is 
\be
\label{symform99}
\begin{split}
[\,{\xi _1}+{\ti \xi _1}\,,\, {\xi _2}+{\ti \xi _2}\,]_{{}_C}
&~=~~ [{\xi _1},{\xi _2}] 
+
\LL_{\ti \xi _1} {\xi _2} - 
\LL_{\ti \xi _2} {\xi _1}
- \tfrac{1}{2}\, \ti d \,(\ti \i_{\ti \xi _1}{\xi _2}-\ti \i_{\ti\xi _2}{\xi _1})\\[1.4ex]
&~~~+[{\ti \xi _1},{\ti \xi _2}] + \LL_{\xi _1}{\ti \xi _2}- \LL_{\xi _2}{\ti \xi _1} -
\tfrac{1}{2}\,d\, (i_{\xi _1}{\ti \xi _2}-i_{\xi _2}{\ti \xi _1}).
\end{split}
\ee
We see that this is precisely the bracket in (\ref{bi-alg}).
For gauge parameters that are independent of $\ti x$, so that 
$\ti \pa =0$
on all quantities, this reduces to 
\begin{align*}
[{\xi _1}+{\ti \xi _1},{\xi _2}+{\ti \xi _2}]_{{}_C}&= [{\xi _1},{\xi _2}] 
+ \LL_{\xi _1}{\ti \xi _2}- \LL_{\xi _2}{\ti \xi _1} -
\tfrac{1}{2}d(i_{\xi _1}{\ti \xi _2}-i_{\xi _2}{\ti \xi _1}),
\end{align*}
which is precisely the Courant bracket~(\ref{cbracket}).

\bigskip

Let us now compute the Jacobiator for the C-bracket.
For this purpose  it is useful to introduce a related
product $\circ$ for 
 fields    
 $P^M, Q^M$
 on $\hat M$, defined by\footnote{For 
 restricted fields, 
the C-bracket becomes the Courant bracket and 
the product
$\circ$ becomes the Dorfman bracket.}
\be
\label{dorfman}  
P \circ Q  \equiv  [ P , Q]  + (\partial ' \hskip-2pt P^M) Q_M \,.
\ee
It then follows (see  (\ref{c-bracket5748})) that the C-bracket and the $\circ$-product differ by a
total derivative:
\be
\label{courant-dorfman99} 
[P, Q]_C  =  P \circ Q  -{1\over 2}\, \partial '\, ( P^MQ_M) \,.
\ee
(Recall 
our notation $\partial ' $ for the derivative $\partial ^M= \eta ^{MN}\partial _N$ with raised index.)
The inner product (\ref{innprod}) is $\langle P , Q \rangle \equiv  P^P Q_P$
and allows us to write
\be
\label{courant-dorfman}  
[P, Q]_C  =  P \circ Q  - {1\over 2}\, \partial ' \langle P, Q \rangle \,.
\ee
While the $\circ$-product is not skew symmetric, its skew-symmetrisation
gives the C-bracket:
\be
[P, Q]_{{}_C} =  {1\over 2} \, (P\circ Q - Q \circ P)\,.
\ee
A key property of the $\circ$-product is that it vanishes when the
first factor is 
 of the form $\partial^M S$, with $S$ a scalar:
\be
\begin{split}
((\partial ' S) \circ  P)^M  
&= [ \partial 'S , P]^M + (\partial^M \partial^K S)P_K \\[0.5ex]
&= (\partial^K S) \partial_K P^M -  P^K \partial_K \partial^MS 
+ (\partial^M \partial^K S)P_K\\[0.5ex]
&=  (\partial^K S) \partial_K P^M  = 0 \,,
\end{split}
\ee
where the 
term in the final line
vanishes by the constraint.  This property,
together with (\ref{courant-dorfman}), gives
\be
\label{coudorf}
\bigl[ \, [ P , Q ]_{{}_C} \,, R \bigr]_{{}_C} =  (P \circ Q) \circ R 
- {1\over 2} \, \partial ' \, \bigl\langle [P, Q]_{{}_C}\,,  R \bigr\rangle\,.
\ee
The $\circ$-product satisfies
a Leibnitz  identity:
\be
P \circ (Q\circ R) = (P \circ Q) \circ R +  Q \circ ( P \circ R) \,.
\ee
This is verified using the definition (\ref{dorfman}) and the Jacobi
identity for the Lie bracket $[ \cdot , \cdot ]$. After some calculation
one is left with a small set of 
terms,  all of which vanish using the
constraint 
that
$(\partial^N X) \partial_N Y =0$ for any $X, Y$.
We now have all the requisite identities.  Following the standard
steps used to compute the Jacobiator for the 
Courant bracket
(see Proposition 3.16 in~\cite{Gualtieri}), we have
\be
\begin{split}
J(P, Q, R) &=  {1\over 4} \Bigl(  P \circ (Q\circ R) + \hbox{c.p.} \Bigr)\\[0.5ex]
&=  {1\over 4} \Bigl( \bigl[ \, [ P , Q ]_{{}_C} \,, R \bigr]_{{}_C}
+ {1\over 2} \, \partial \, \bigl\langle [P, Q]_{{}_C}\,,  R \bigr\rangle  + \hbox{c.p.} \Bigr)\,.   \end{split}
\ee
Here c.p. stands for cyclic permutation. We thus get
\be
J(P, Q, R) =  {1\over 6}\,  \partial ' \Bigl(  
 \bigl\langle [P, Q]_{{}_C}\,,  R \bigr\rangle  + \hbox{c.p.}
  \Bigr)\,.   
\ee
This is of course  consistent with the Jacobiator for the Courant bracket given earlier.

\sectiono{Conclusions}

In this paper we have focused on fields restricted to some null  subspace $N$ of 
the doubled space~$\hat M$ and 
investigated the consequences of this restriction for  the theory  of~\cite{Hull:2009mi}.
 Since the space $N$ need not be
specified, the field theory has $O(D,D)$ covariance.  
The results of Siegel~\cite{Siegel:1993th} also
give an  $O(D,D)$ covariant field theory defined on such an $N$.  
While the theory of~\cite{Siegel:1993th} involved geometric fields on $\hat M$ with additional gauge invariances, ours involves just  the fields that  arise in closed string field theory. It will be interesting to compare the two approaches further, and in particular the consequences of the failure of the Jacobi identities.
The restriction to $N$  implies that these
field theories do {\em not} really include both momentum and winding, as they are T-dual to ones without winding. 

For the full gauge algebra of double field theory we are interested in the C-bracket for arbitrary 
gauge parameters on $\hat M$  that satisfy the $\D =0$ 
constraint but are not necessarily restricted to some null space $N$. The C-bracket of such fields takes the elegant and symmetric form 
(\ref{symform99}), with a projector to $\D =0$ implicit, as in~\cite{Hull:2009mi}. It would be interesting to investigate the structure of the Jacobiator in this general setup. The results in this paper and the understanding
gained on the relation of the C-bracket to  the Courant bracket should
play an important role in 
the
construction of  the full double field theory for fields not restricted to any null subspace.

Our gauge algebra is given by the C-bracket on the doubled space 
$\hat M$.  This algebra 
 is distinct from the diffeomorphism algebra on  
$\hat M$ which would be given by the Lie bracket.
A striking feature is that, when restricted to {\it any} $N$,  the gauge parameters
$\x^M$ decompose into a vector field and 1-form on $N$ 
that are parameters for diffeomorphisms and 
$B$-field gauge transformations.
 As discussed at the end of \S\ref{dilscavec},  
 our gauge parameters $\x^M$ are not 
 conventional vector fields on $\hat M$ but are instead C-vectors transforming as in  (\ref{c-vector})  and identified under the transformations $\x^M\to \x^M +\pa^M\chi$.
Although we have diffeomorphism symmetry on every $N$, 
the  lift to the doubled space $\hat M$  
 gives symmetries 
 with a C-bracket algebra
 that appear to be distinct from diffeomorphisms  on $\hat M$ and which will generalise to unrestricted fields.
 It will be interesting to understand the symmetry structure arising from the C-bracket and the generalisation to unrestricted fields further.  We hope to return to these issues in the future.

\vspace{0.6cm}

{\bf \large Acknowledgments:}  
We are grateful to Marco 
Gualtieri for several instructive exchanges on the subject of
Courant brackets  
and to Dan Waldram for helpful discussions.
B.Z. is supported in part by the U.S.
DOE grant De-FC02-94eR40818.

\appendix

\sectiono{Commutator of gauge transformations}
\setcounter{equation}{0}

We use matrix notation to rewrite the $e_{ij}$ transformation
(\ref{finalgt}) as 
$\delta_\lambda e= \delta_\lambda^{(0)}e
+ \delta_\lambda^{(1)}e+ \delta_\lambda^{(2)}e$, 
with 
\be
\begin{split}
\delta_\lambda^{(0)}e&= ~M(\lambda, \bar \lambda)  \,,  \\[0.5ex]
\delta_\lambda^{(1)}e& = ~{\cal O}(\lambda, \bar \lambda) \,e + 
N(\lambda)\,  e\, - e \,\bar N (\bar \lambda) \,,\\
\delta_\lambda^{(2)} e& =  - {1\over 4}  \,e  M^t(\lambda, \bar \lambda)\, e \,.
\end{split}
\ee
In here we have introduced the differential operator 
${\cal O}$ and the matrices $M, N,$ and $\bar N$ defined by
\be
\begin{split}  
{\cal O} (\lambda, \bar \lambda) &\equiv 
 {1\over 2}\,(\lambda \cdot D + \bar \lambda \cdot \bar D)\,, \\[1.0ex]
 M_{ij}(\lambda, \bar \lambda) &\equiv D_i\bar\lambda_j  + \bar D_j 
\lambda_i  \,,\\[1.0ex]
 N_{ij} (\lambda) &\equiv  D_i \lambda^j   -  D^j \lambda_i \,,\\[1.0ex]
 \bar  N_{ij} (\bar \lambda) &\equiv  \bar D^i \bar \lambda_j   -  \bar D_j 
 \bar\lambda^i\,.
\end{split}
\ee
The gauge algebra requires
$
\bigl[ \, \delta_{\lambda_1} \,, \,  \delta_{\lambda_2} \, \bigr] = \delta_\Lambda
$ and this condition results in
\be
\label{kssvm}
\begin{split}
\delta_\Lambda^{(0)} &=
\bigl[\,\delta_{\lambda_1}^{(0)} 
\,,\,\delta_{\lambda_2}^{(1)} \, \bigr]\,, \\[1.0ex]
\delta_\Lambda^{(1)} &=
\bigl[ \, \delta_{\lambda_1}^{(1)} \,, \,  \delta_{\lambda_2}^{(1)} \, \bigr]
+\bigl[ \, \delta_{\lambda_1}^{(0)} \,, \,  \delta_{\lambda_2}^{(2)} \, \bigr]
+\bigl[ \, \delta_{\lambda_1}^{(2)} \,, \,  \delta_{\lambda_2}^{(0)} \, \bigr]
\,,\\[1.0ex]
\delta_\Lambda^{(2)} &=
\bigl[ \, \delta_{\lambda_1}^{(1)} \,, \,  \delta_{\lambda_2}^{(2)} \, \bigr]
+\bigl[ \, \delta_{\lambda_1}^{(2)} \,, \,  \delta_{\lambda_2}^{(1)} \, \bigr]
\,,\\[1.0ex]
0~~ &=
\bigl[ \, \delta_{\lambda_1}^{(2)} \,, \,  \delta_{\lambda_2}^{(2)} \, \bigr]\,.
\end{split}
\ee
The first condition requires
\be
\label{ksssg}
M(\Lambda, \bar\Lambda)\, = \, {\cal O}(\lambda_2, \bar\lambda_2) M (\lambda_1, \bar\lambda_1) + N(\lambda_2) M(\lambda_1) 
+ M(\lambda_2, \bar \lambda_2) \bar N (\bar\lambda_1) \, 
-\,  \{\lambda_1\leftrightarrow \lambda_2\}\,.
\ee
We have checked this equation works out correctly.  The second
condition in (\ref{kssvm}), for the algebra to hold with terms linear
on the fields, requires the following conditions:
\be
\label{ksscl}
\begin{split}
{\cal O}(\Lambda, \bar \Lambda)  &=   \bigl[ {\cal O} (\lambda_2, \bar\lambda_2) \,, \,  {\cal O} (\lambda_1, \bar\lambda_1) \, \bigr] \,,\\[1.0ex]
N(\Lambda)  &=    {\cal O} (\lambda_2, \bar\lambda_2) N(\lambda_1)  
+ N(\lambda_2) N(\lambda_1) + {1\over 4} 
\, M(\lambda_2, \bar\lambda_2)
  M^t(\lambda_1, \bar\lambda_1) 
  -\,  \{\lambda_1\leftrightarrow \lambda_2\} \,, \\[1.0ex]
\bar N(\bar \Lambda)  &=    {\cal O} (\lambda_2, \bar\lambda_2) 
\bar N(\bar\lambda_1)  
+ \bar N(\bar\lambda_2) \bar N(\bar \lambda_1) + {1\over 4} 
\, M^t(\lambda_2, \bar\lambda_2)
  M(\lambda_1, \bar\lambda_1) 
  -\,  \{\lambda_1\leftrightarrow \lambda_2\} \,.
\end{split}
\ee
The first equation 
is straightforward to
establish.  
The second equation can be proven with a bit of algebra.
The contribution from the $M M^t$ terms
is needed  
to get the identity to work, confirming that the $[\delta^{(2)},
\delta^{(0)}]$ commutator is crucial to get the gauge algebra to close
at this order.  The last
equation  in (\ref{ksscl})
is  a consequence of the second
and the following discrete symmetry.  As we let $D_i\to \bar D_i, ~ \bar D_j \to D_j, ~\lambda_i \to \bar \lambda_i,$ and  $\bar \lambda_j \to \lambda_j$ we find that ${\cal O}(\lambda, \bar\lambda)$ is invariant and 
\be
\Lambda_i \to \bar\Lambda_i\, , ~~ \bar \Lambda_j \to \Lambda_j \,,
~~  N (\lambda) \to \bar N(\bar \lambda)\,, ~~ M (\lambda, \bar \lambda)
\to  M^t (\lambda, \bar \lambda) \,.
\ee
The third condition in (\ref{kssvm}) guarantees that the terms quadratic
in fields work out.  A little calculation shows that this condition 
requires
\be
\label{ksssg99}  
M^t(\Lambda, \bar\Lambda)\, = \, {\cal O}(\lambda_2, \bar\lambda_2) 
M^t (\lambda_1, \bar\lambda_1) + \bar N(\bar\lambda_2) M^t(\lambda_1) 
+ M^t(\lambda_2, \bar \lambda_2) N (\lambda_1) \, 
-\,  \{\lambda_1\leftrightarrow \lambda_2\}  \,.   
\ee
This actually holds on account of (\ref{ksssg}) and the discrete symmetry.
The last equation in (\ref{kssvm}) is needed for the commutator algebra
not to acquire cubic terms in the fields.  It is quickly confirmed.  
This concludes our verification of the algebra.

\end{document}